\documentclass[10pt,tightenlines,preprintnumbers,showpacs,
superscriptaddress,nofootinbib]{revtex4}
\newcommand{\PRE}[1]{{#1}} 
\usepackage{bm}
\usepackage{subfigure}
\usepackage{slashed}
\usepackage{epsfig}
\usepackage{hyperref}

\def\beq{\begin{eqnarray}}
\def\eeq{\end{eqnarray}}
\def\bea{\begin{eqnarray}}
\def\eea{\end{eqnarray}}

\newcommand{\gev}{\text{GeV}}

\newcommand{\eqref}[1]{Eq.~(\ref{#1})}

\newcommand{\figref}[1]{Fig.~\ref{fig:#1}}
\newcommand{\Figref}[1]{Figure~\ref{fig:#1}}

\newcommand{\gsim}{\lower.7ex\hbox{$\;\stackrel{\textstyle>}{\sim}\;$}}
\newcommand{\lsim}{\lower.7ex\hbox{$\;\stackrel{\textstyle<}{\sim}\;$}}

\newcommand{\ssection}[1]{ \vspace*{-0.05in} \section{#1} \vspace*{-0.08in}}
\newcommand{\ssectionstar}[1]{ \vspace*{-0.05in} \section*{#1} \vspace*{-0.08in}}

\begin{document}

\preprint{UCI-TR-2013-15, UH511-1214-2013, CALT 68-2944, CETUP2013-001}

\title{ \PRE{\vspace*{0.0in}}
Isospin-Violating Dark Matter Benchmarks for Snowmass 2013
\PRE{\vspace*{0.1in}} }

\author{Jonathan L.~Feng}
\affiliation{Department of Physics and Astronomy, University of
California, Irvine, CA 92697, USA
}

\author{Jason Kumar}
\affiliation{Department of Physics and Astronomy, University of
Hawaii, Honolulu, HI 96822, USA
}

\author{Danny Marfatia}
\affiliation{Department of Physics and Astronomy,
University of Kansas, Lawrence, KS 66045, USA
}
\affiliation{Kavli Institute for Theoretical Physics, University of
  California, Santa Barbara, CA 93106, USA}

\author{David Sanford}
\affiliation{Department of Physics, California Institute of
  Technology, Pasadena, CA 91125, USA
}

\begin{abstract}
\PRE{\vspace*{.0in}} Isospin-violating dark matter (IVDM) generalizes
the standard spin-independent scattering parameter space by
introducing one additional parameter, the neutron-to-proton coupling
ratio $f_n/f_p$.  In IVDM the implications of direct detection
experiments can be altered significantly.  We review the motivations
for considering IVDM and present benchmark models that illustrate some
of the qualitatively different possibilities.  IVDM strongly motivates
the use of a variety of target nuclei in direct detection experiments.
\end{abstract}


\maketitle

\ssection{Introduction}

The standard presentation of direct detection experimental results for
spin-independent scattering is in the $(m_X, \sigma_p)$ plane, where
$m_X$ is the mass of the dark matter particle $X$, and $\sigma_p$ is
the $X$-proton scattering cross section.  However, direct detection
experiments do not directly constrain $\sigma_p$. Rather, they bound
scattering cross sections off of nuclei.  Results for nuclei are then
interpreted as bounds on $\sigma_p$ by assuming that the couplings of
dark matter to protons and neutrons are identical, {\it i.e.}, that
the dark matter's couplings are isospin-invariant.

This assumption is valid if the interaction between dark matter and
quarks is mediated by a Higgs boson, as in the case of neutralinos
with heavy squarks. In general, however, it is not theoretically well-motivated:
the assumption is violated by many dark matter candidates,
including neutralinos with light squarks, dark matter with
$Z$-mediated interactions with the standard model (SM), dark matter
charged under a hidden U(1) gauge group with a small kinetic mixing
with hypercharge, and dark matter coupled through new scalar or
fermionic mediators with arbitrary flavor structure.  In the next ten
to twenty years, during which many direct detection experiments will
be searching for dark matter, it is clearly desirable to consider
frameworks that can accommodate these more general possibilities.

Isospin-violating dark matter (IVDM)~\cite{Kurylov:2003ra,
  Giuliani:2005my,Chang:2010yk,Kang:2010mh,Feng:2011vu,Feng:2013vod}
provides a simple framework that accommodates all these possibilities
by including a single new parameter, the neutron-to-proton coupling
ratio $f_n/f_p$.  One might have expected an overarching framework to
need many more parameters. However, for spin-independent scattering
with the typical energies of weakly-interacting massive particle
(WIMP) collisions, the dark matter does not probe the internal
structure of nucleons.  Loosely speaking, dark matter sees nucleons,
but not quarks.  Nucleons are therefore the correct ``effective
degrees of freedom'' for spin-independent WIMP scattering, and IVDM
therefore captures all of the possible variations by letting the
proton couplings differ from the neutron couplings.

Although a simple generalization, IVDM can drastically change the
interpretation of data from direct detection experiments.  This aspect
has been highlighted with respect to data at low mass ($5-20~\gev$),
in which several potential signals have been reported
(DAMA~\cite{Bernabei:2010mq},
CoGeNT~\cite{Aalseth:2010vx,Aalseth:2012if},
CRESST~\cite{Angloher:2011uu}, and CDMS-Si~\cite{Agnese:2013rvf}) and
several bounds have been placed
(CDMS-Ge~\cite{Akerib:2010pv,Ahmed:2010wy},
Edelweiss~\cite{Armengaud:2012pfa}, XENON10~\cite{Angle:2011th}, and
XENON100~\cite{Aprile:2011hi,Aprile:2012nq}).  With the assumption of
isospin invariance, many of the signal regions of interest (ROIs) do
not overlap, and almost all of the ROIs are excluded by null results
from other experiments.  The assumption of isospin invariance is
especially unmotivated for low-mass dark matter since isospin
invariance is primarily motivated by neutralino dark matter, which
cannot explain the low-mass data in standard supersymmetric
frameworks.  Although IVDM does not make it possible to reconcile all
of the existing data at present, it can alter the standard picture
drastically, and its implications for low-mass dark matter, although
not the primary reason to consider IVDM, illustrate well how different
the sensitivities of various experiments may be once the assumption of
isospin invariance is relaxed.

\ssection{Formalism}

Dark matter-nuclei scattering is largely coherent, which for
isospin-invariant scenarios produces a well-known $A^2$ enhancement to
the cross section, favoring scattering off heavier elements.  But in
the case of isospin-violation, destructive interference can instead
suppress the scattering cross section.  Although direct detection
experiments typically present results in terms of $\sigma_p$, the
actual quantity reported is the {\em normalized-to-nucleon cross
  section} $\sigma_N^Z$, which is the dark matter-nucleon scattering
cross section that is inferred from the data of a detector with a
target with $Z$ protons, assuming isospin-invariant interactions.
This quantity is related to $\sigma_p$ by the ``degradation
factor''~\cite{Feng:2013vod}
\begin{eqnarray}
D^Z_p \equiv \frac{\sigma_N^Z}{\sigma_p} &=& \frac{\sum_i \eta_i
  \mu_{A_i}^2
[Z + (f_n / f_p) (A_i -Z)]^2 }{\sum_i \eta_i \mu_{A_i}^2 A_i^2} \, ,
\label{DZ}
\end{eqnarray}
where $\eta_i$ is the natural abundance of the $i^{\rm {th}}$ isotope,
$\mu_{A_i} = m_X m_{A_i} /(m_X + m_{A_i})$ is the reduced mass of the dark
matter-nucleus system, and $f_n$ and $f_p$ are the couplings of dark
matter to neutrons and protons, respectively.  For isospin-invariant
interactions, $f_n = f_p$, and $\sigma_N^Z = \sigma_p$.

Although $\sigma_p$ is not directly measured, a determination of the
normalized-to-nucleon cross section by two detectors with different
targets provides a measurement of $\sigma_N^{Z_1} / \sigma_N^{Z_2} =
D_p^{Z_1} /D_p^{Z_2}$.  From \eqref{DZ}, this quantity depends
quadratically on $f_n / f_p$.  Measurements of the
normalized-to-nucleon cross section by two experiments with different
targets are thus sufficient to determine $f_n / f_p$ up to a two-fold
ambiguity.  A measurement with a third target material is required to
break this degeneracy.

\ssection{Benchmarks}

Absent any prejudice, $f_n / f_p$ is a free parameter that must be
constrained by data, no different than the mass and cross section.
But we can identify some benchmark values of $f_n / f_p$ that are
particularly noteworthy:
\begin{enumerate}
\item $f_n / f_p = -13.3$ (``$Z$-mediated''): Valid for dark
  matter with $Z$-mediated interactions with the SM.
\item $f_n / f_p = -0.82$ (``Argophobic''): For this value, the sensitivity
  of argon-based detectors is maximally degraded.  Note that potential
  CoGeNT and CDMS-Si signals can be made consistent for
  $f_n /f _p = -0.89$~\cite{Feng:2013vod}. (The other region for which these
  signals can be consistent includes the isospin-invariant case.)
\item $f_n / f_p = -0.70$ (``Xenophobic''): For this value, the
  sensitivity of xenon-based detectors is maximally degraded.
\item $f_n / f_p =0$ (``Dark photon-mediated''): Valid for dark
  matter that interacts with the SM through kinetic mixing with the
  photon.
\item $f_n / f_p = 1$ (``Isospin-invariant''): Valid for dark
  matter that interacts with the SM through Higgs exchange.
\end{enumerate}

\ssection{Impact on Direct Detection}

\begin{figure}[tb]
\subfigure[\ Entire $f_n/f_p$ range]{
\includegraphics*[width=0.47\textwidth]{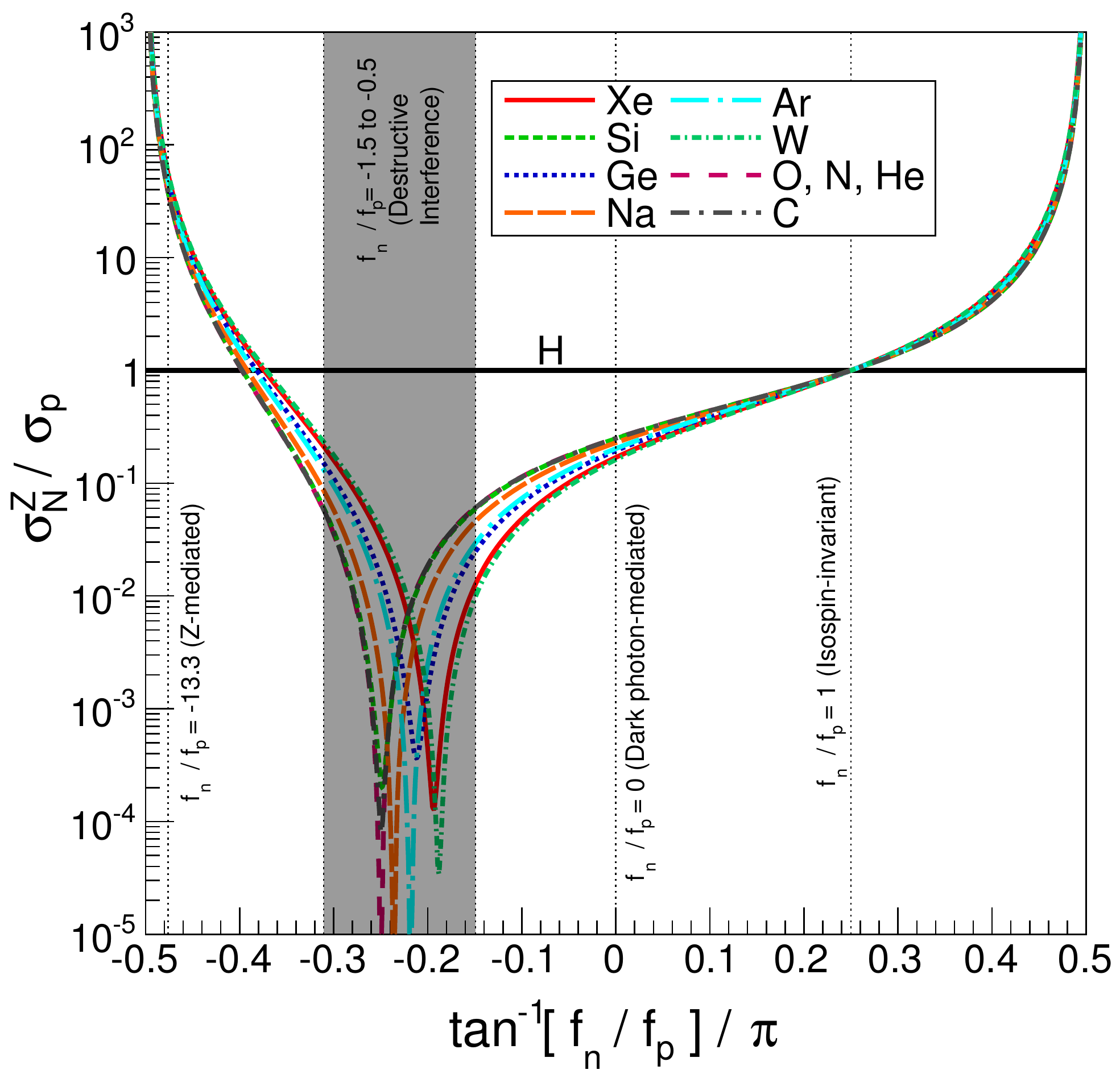}
\label{fig:FZp}}
\subfigure[\ Destructive interference region]{
\includegraphics*[width=0.47\textwidth]{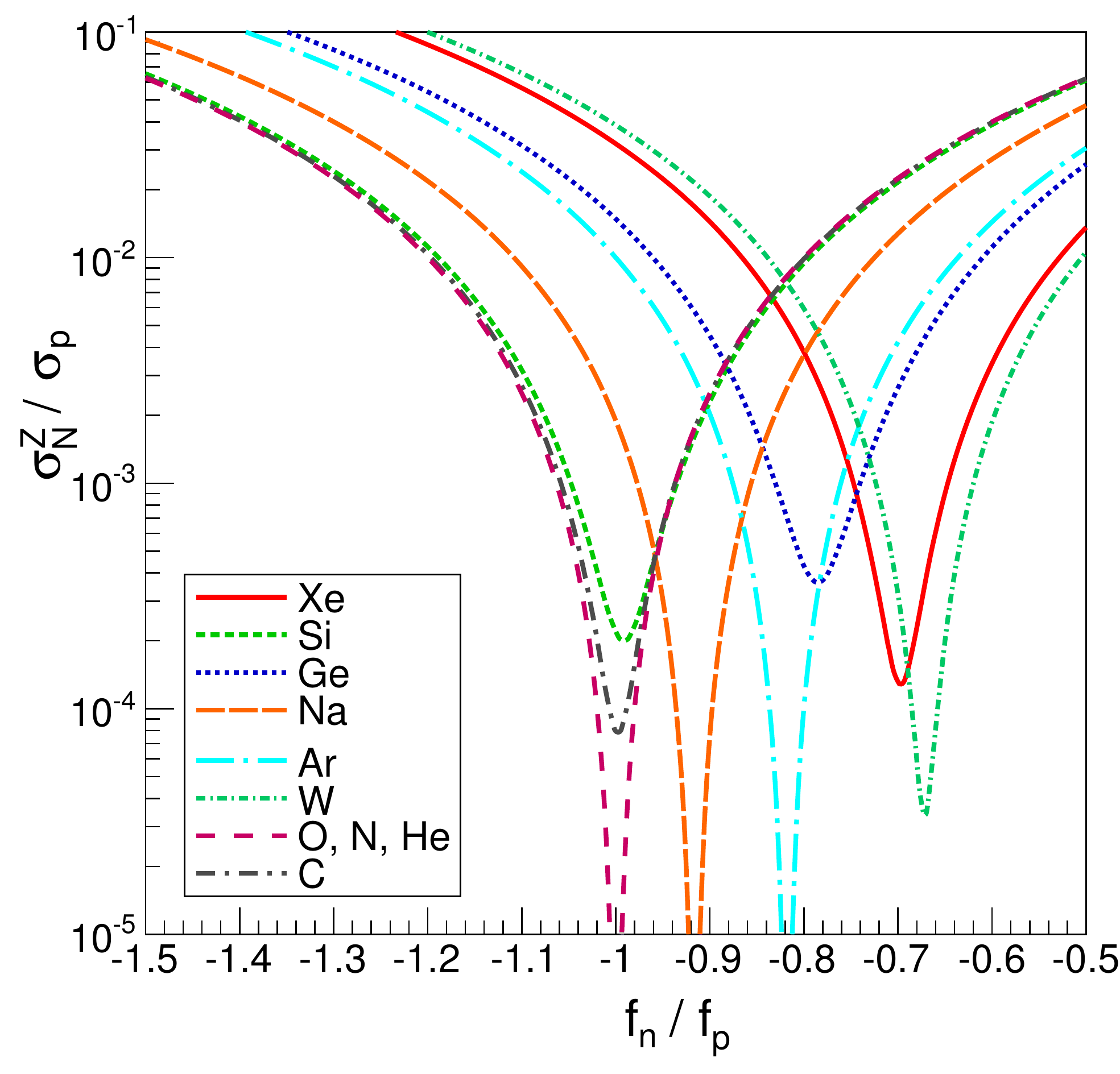}
\label{fig:FZp_small}}
\vspace*{-.1in}
\caption{\label{fig:FZ} Ratio of $\sigma^N_Z$ to $\sigma_p$ for
  materials relevant to direct detection
  experiments~\cite{Feng:2013vod}.  Ratios are shown as a function of
  $f_n / f_p$ for (a) the entire range of couplings and (b) the
  destructive interference region.
  We have made the mild assumption that the reduced masses $\mu_{A_i}$ are all equal for a given element and dark matter mass.
}
\end{figure}

In \figref{FZ} we plot $\sigma_N^Z / \sigma_p$ as a function of $f_n /
f_p$ for many of the target materials commonly used for direct
detection experiments.  The full range of $f_n / f_p$ is shown in
\figref{FZp} and the destructive interference region
($-1.5 \leq f_n / f_p \leq -0.5$) is
shown in \figref{FZp_small}.  For materials with only one isotope with
significant abundance, such as oxygen, nitrogen, helium, sodium, and
argon, it is possible to almost completely eliminate the detector's
response with a particular choice of $f_n / f_p$.  But for a material
such as xenon, with many isotopes, it is not possible to cancel the
response of all isotopes simultaneously.  For materials such as
carbon, silicon, germanium, xenon, and tungsten, the maximum factor by
which their sensitivity to $\sigma_p$ may be degraded is within the
range $10^{-5} - 10^{-3}$.

\begin{figure}[tb]
\subfigure[]{ \includegraphics*[width=0.47\textwidth]{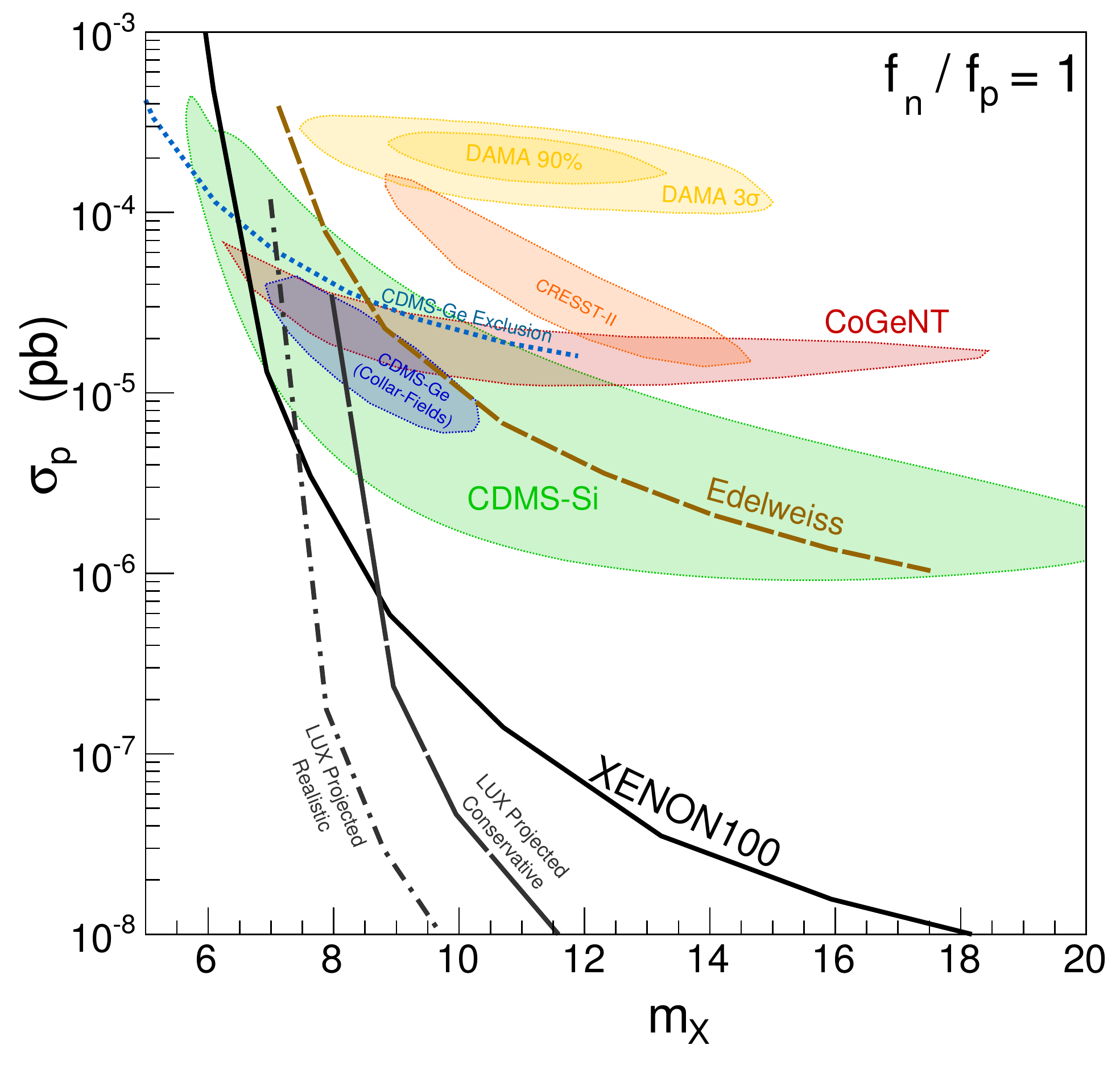}
\label{fig:sigmap_1}}
\subfigure[]{
\includegraphics*[width=0.47\textwidth]{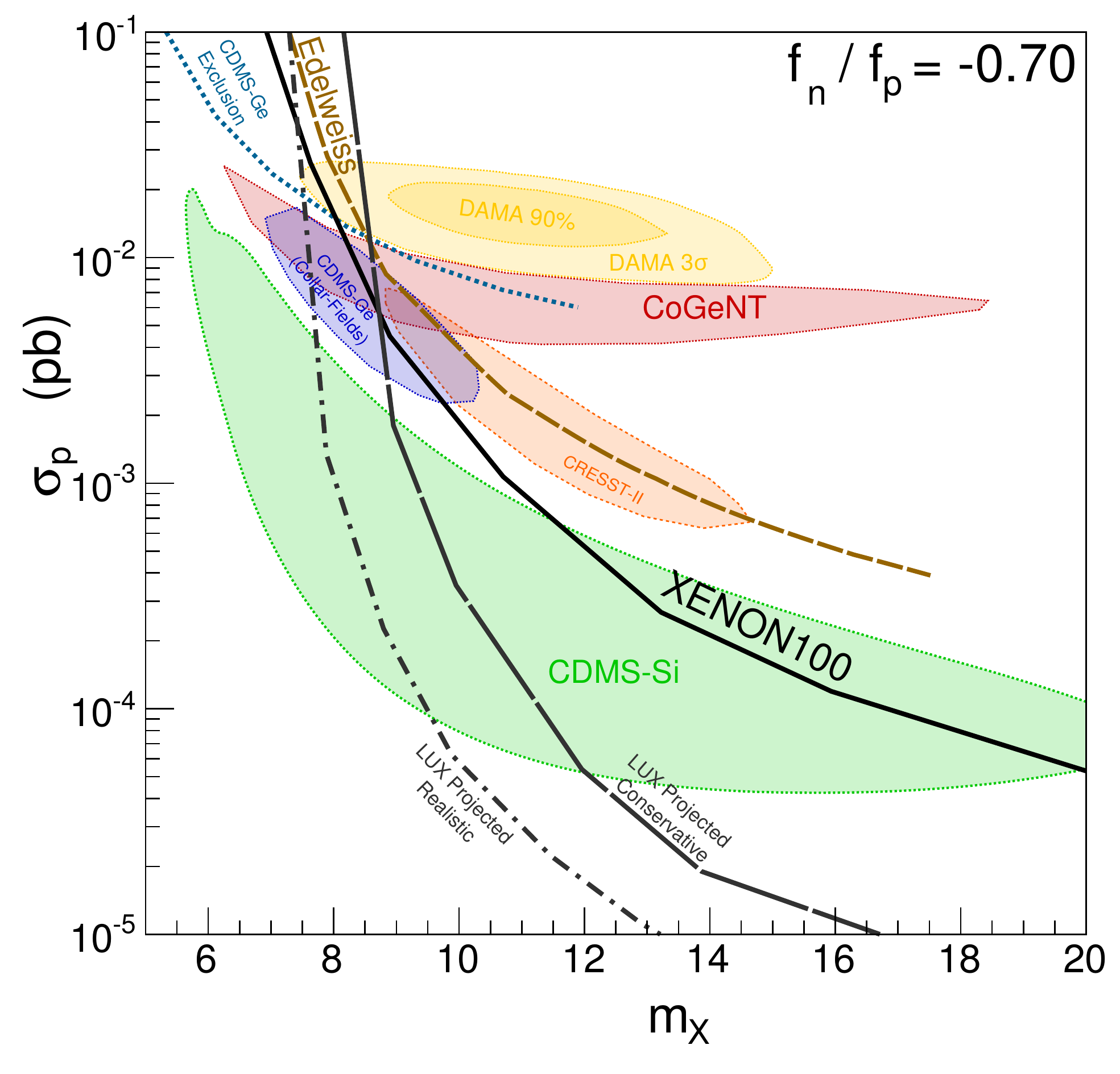}
\label{fig:sigmap_-7}}
\vspace*{-.1in}
\caption{\label{fig:sigmap_onepoint} Light dark matter experimental
  results in the $(m_X, \sigma_p)$ plane for (a) the isospin-invariant
  case $f_n / f_p = 1$ and (b) the xenophobic case $f_n / f_p =
  -0.70$~\cite{Feng:2013vod}. Plotted are 90\% CL ROIs for
  CoGeNT~\cite{Aalseth:2010vx}, CRESST~\cite{Angloher:2011uu},
  CDMS-Si~\cite{Agnese:2013rvf}, an ROI for an independent analysis of
  CDMS-Ge data~\cite{Collar:2012ed}, the 90\% and $3\sigma$ ROIs for
  DAMA~\cite{Bernabei:2010mq} as determined in
  Refs.~\cite{Savage:2008er, Savage:2010tg}.  Exclusion contours from
  CDMS~\cite{Ahmed:2010wy}, Edelweiss~\cite{Armengaud:2012pfa}, and
  XENON100~\cite{Aprile:2011hi,Aprile:2012nq} are also shown, as are
  projected bounds from LUX~\cite{Akerib:2012ak}.}
\end{figure}

\Figref{sigmap_onepoint} shows relevant direct detection constraints
and possible signals in the dark matter mass range $5-20~\gev$.  For
the isospin-invariant case shown in \figref{sigmap_1}, $f_n / f_p =
1$, XENON100 results~\cite{Aprile:2012nq} place stringent constraints
on the parameter space.  On the other hand, for the xenophobic value
$f_n / f_p = - 0.70$ shown in \figref{sigmap_-7}, the CDMS-Si ROI
almost entirely evades the XENON100 bound, and the ROIs from
CoGeNT~\cite{Aalseth:2010vx} and an ROI from an independent reanalysis
of CDMS-Ge data~\cite{Collar:2012ed} become marginally consistent with
the XENON100 bound.  However, the DAMA~\cite{Bernabei:2010mq} and
CRESST~\cite{Angloher:2011uu} ROIs remain in tension with XENON100
bounds for $f_n/f_p = -0.70$, and the agreement between CDMS-Si and
the CoGeNT and CDMS-Ge results is weakened.

\ssection{Complementary Astrophysical and Collider Probes}

IVDM models can also be probed through monojet/monophoton collider
searches~\cite{Rajaraman:2011wf,Kumar:2011dr,Hagiwara:2012we,Feng:2013vod} and
indirect detection searches using the galactic center, galactic halo, dwarf
spheroidals, etc.~as sources~\cite{Evoli:2011id,Kumar:2011dr,Jin:2012jn}.
To compare sensitivities, one typically considers a particular dark
matter-parton interaction structure that generates spin-independent
scattering and reproduces direct detection data for a particular
choice of $f_n/ f_p$.  Crossing symmetry is then used to determine the
dark matter annihilation or collider production cross section.

For IVDM with destructive interference, to maintain the same direct
detection cross sections, the individual couplings to first generation
quarks must be enhanced, which implies enhanced cross sections for
dark matter annihilation and dark matter production at colliders.
Moreover, both indirect and collider searches tend to have greater
sensitivity to low-mass dark matter.  At the same time, it is
important to note that indirect detection bounds are weakened if the
dark matter-parton interaction structure permits only $P$-wave
annihilation, and both collider and indirect detection sensitivities
are weakened if dark matter couples to a light mediator.

Other interesting complementary probes of IVDM arise from searches for
neutrinos arising from dark matter annihilation in the
sun~\cite{Kumar:2011hi,Chen:2011vda,Gao:2011bq}.  If the dark matter
scattering and annihilation processes in the sun are in equilibrium,
the neutrino event rate is directly related to the dark matter solar
capture rate, which in turn is proportional to the cross section for
dark matter to scatter off solar nuclei.  Since the sun is dominated
by elements with small numbers of neutrons, it provides targets that
are complementary to targets like germanium and xenon.

\ssection{Conclusions}

The main motivation for isospin-violating dark matter is theoretical.
Dark matter with a mass $\agt 1~\gev$ does not probe the internal
structure of nucleons, but does probe the nucleon structure of nuclei;
a framework that treats the dark matter coupling to protons and
neutrons as independent parameters is thus the most natural framework
for describing dark matter interactions.  Isospin-violating
interactions can have a large impact on the way direct detection data
is interpreted, potentially helping to reconcile some of the seemingly
inconsistent data from direct detection experiments at low mass.  A
complete determination of the isospin structure of dark matter
interactions would require data from at least three direct detection
experiments with different targets.  However, data from indirect
detection or collider searches can potentially provide complementary
data that can help determine $f_n / f_p$, especially at low mass.

\ssectionstar{Acknowledgments}

We are grateful to T.~Cohen and D.~McKinsey for useful discussions.
JLF is supported in part by U.S.~NSF grant No.~PHY--0970173 and by the
Simons Foundation.  JK is supported in part by U.S.~NSF CAREER Award
PHY-1250573. DM is supported by the U.S.~DOE under Grant
No. DE--FG02--13ER42024.  DS is supported in part by U.S.~DOE grant
DE--FG02--92ER40701 and by the Gordon and Betty Moore Foundation
through Grant No.~776 to the Caltech Moore Center for Theoretical
Cosmology and Physics.  JK and DS thank the Center for Theoretical
Underground Physics and Related Areas (CETUP* 2013) in South Dakota
for its support and hospitality, and DM thanks the Kavli Institute for
Theoretical Physics for its support (via U.S.~NSF grant
No.~PHY11--25915) and hospitality during the completion of this work.

\bibliography{ivdmwhitepaperArXiv2}{}

\end{document}